\documentclass{iopart}
\usepackage{graphicx}

\begin{document}

\title[Macromolecular tracer diffusion in polymer solutions]{A fluorescence 
correlation spectroscopy study of macromolecular tracer diffusion in polymer 
solutions}

\author{Ute Zettl$^1$, Matthias Ballauff$^{\hspace*{0.05cm}2}$, Ludger Harnau$^3$}

\address{$^1$Physikalische Chemie I, University of Bayreuth, 
95440 Bayreuth, Germany\\
$^2$Soft Matter and Functional Materials, Helmholtz-Zentrum Berlin,
14109 Berlin, Germany\\
$^3$Max-Planck-Institut f\"ur Metallforschung, Heisenbergstr. 3, 
70569 Stuttgart, Germany, 
and Institut f\"ur Theoretische und Angewandte Physik, Universit\"at Stuttgart, 
Pfaffenwaldring 57, 70569 Stuttgart, Germany}
\ead{harnau@fluids.mpi-stuttgart.mpg.de}

\begin{abstract}
We discuss the manner in which the dynamics of tracer 
polystyrene chains varies with the concentration of matrix polystyrene chains 
dissolved in toluene. Using fluorescence correlation spectroscopy and 
theory, it is shown that the cooperative diffusion coefficient of the matrix
polystyrene chains can be measured by fluorescence correlation spectroscopy 
in the semidilute entangled concentration regime. In addition the 
self-diffusion coefficient of the tracer polystyrene chains can be detected 
for arbitrary concentrations. The measured cooperative diffusion coefficient 
is independent of the molecular weight of the tracer polystyrene chains 
because it is a characteristic feature of the transient entanglement network.
\end{abstract}
\maketitle

\section{Introduction}
The dynamics of solutions composed of high molecular weight polymer chains 
differs qualitatively from the dynamics of simple fluids due to entanglements.
At a microscopic scale, entanglements arise from the fact that linear 
polymer chains are one-dimensionally connected objects which cannot cross 
each other. The resulting topological interaction strongly affects dynamical 
properties since it imposes constraints on the motion of the polymer chains.
The common interpretation of the physical origin of entanglement phenomena 
is that a transient network emerges due to the interactions between the 
polymer chains. The so-called reptation model is the most developed and 
widely applied phenomenological theory for the motion of polymer chains
(see e.g., ref \cite{doi:86} and references therein). It focusses on the 
motion of a single polymer chain in an static field due to the surrounding 
polymer chains. Although widely accepted, there are significant discrepancies 
between predictions of the reptation model and experiments on polymer solutions 
because the surrounding polymer chains due not act as a static field but 
exhibit themselves as cooperative fluctuations. Hence there is an important 
coupling between the single polymer chain motion and the cooperative network 
fluctuations. As a result of this coupling it should be possible to detect 
both the single polymer and the cooperative dynamics using the same experimental 
technique.

\begin{figure}[ht!]
\begin{center}
\includegraphics[width=6cm, clip]{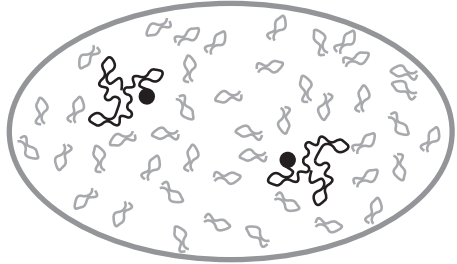}
\caption{Schematic illustration of a polymer solution composed of a small 
concentration of labeled tracer polymer chains (black wriggled lines) and 
matrix polymer chains of different molecular weight (gray wriggled lines).
Each labeled polymer chain carries only one dye molecule at one of its ends 
which is marked by a black dot. The FCS observation volume is enclosed by the 
thick gray ellipsoidal line. The size of the polymer chains, the size 
observation volume, and the number of polymer chains are not drawn to absolute 
scale. Only the fact that the molecular weight of the matrix polymer chains 
can be different from that of the tracer polymer chains is relevant.}
\label{fig1}
\end{center}
\end{figure}

Recently it has been demonstrated that two different diffusion coefficients 
can be obtained with fluorescence correlation spectroscopy (FCS) using
single-labeled polystyrene (PS) solutions \cite{zett:09}.
The self-diffusion coefficient $D_s(c)$ results from FCS in the limit of small
concentrations of labeled PS chains and for arbitrary concentrations $c$ of 
unlabeled PS chains. Moreover, the cooperative 
diffusion coefficient $D_{co}(c)$ becomes accessible in the semidilute entangled 
regime due to an  effective long-range interaction in the transient entanglement
network. The self-diffusion coefficient describes the motion of one molecule 
relative to the surrounding molecules due to thermal motions while the cooperative
diffusion coefficient describes the dynamics of a number of molecules in a 
concentration gradient. It has been pointed out that measurements of both 
$D_s(c)$ and $D_{co}(c)$ are very interesting since a central problem in the dynamics 
of semidilute entangled polymer solutions is the quantitative understanding of 
the interplay of self-diffusion and cooperative diffusion. Motivated by this 
prospect, we therefore extend our previous study 
\cite{zett:09} to the case of macromolecular tracer diffusion as is illustrated 
in figure \ref{fig1}. Hence we study the dynamics of long end-labeled tracer PS 
chains (black wriggled lines in figure \ref{fig1}) immersed in a polymer solution 
consisting of shorter matrix PS chains (gray wriggled lines in figure \ref{fig1}).
The study is devoted to an understanding of the coupling of self- and 
cooperative motion due to topological constraints. Varying the concentration 
and the molecular weight of the matrix PS chains allows us to modify these topological 
constraints.
%
%
%

\section{Dye labeled tracer polystyrene chains}
Linear PS chains with different molecular weights and low polydispersity 
were prepared as discussed in ref. \cite{zett:04}. Our samples were composed of 
a small concentration (around $10^{-8}\;\mathrm{M}$) of Rhodamine~B labeled tracer 
PS chains of molecular weight $M_{w}^{(tr)}$ in toluene solutions in which 
matrix PS chains of molecular weight $M_{w}^{(ma)}$  were dissolved at various 
concentrations. The resulting systems are denoted as \mbox{PS-$M_{w}^{(tr)}$/$M_{w}^{(ma)}$} 
as is indicated in table \ref{tab1}. Each labeled PS chain carries only one dye molecule 
at one of its ends. We have used preparative gel permeation chromatography to separate 
labeled polymer chains and free dye molecules \cite{zett:04,zett:05}. Therefore, 
the resulting PS solutions do not contain any measurable amount of free dye molecules.
%
%
\begin{table}[ht!]
\caption{\label{tab1} Molecular weights $M_{w}^{(tr)}$ and $M_{w}^{(ma)}$ of the 
tracer and matrix PS chains, respectively. The concentration at which 
the fast diffusion process appears in the FCS measurements is denoted as $c^+$. 
In the main text and the figures the various PS solutions are denoted by the names 
given in the first column of the table. For comparison the overlap concentration 
$c^\star$ of the matrix PS chains is shown in the last column. \cite{zett:07}}

\begin{indented}
\lineup
\item[]\begin{tabular}{@{}*{5}{l}}
\br
Name & $M_{w}^{(tr)}$ [kg/mol] & $M_{w}^{(ma)}$ [kg/mol] 
& $c^+[\mathrm{wt\%}]$ & $c^\star[\mathrm{wt\%}]$\cr\mr
PS-17/17    &  17  & 17  &        &  3.5  $\pm$ 0.3 \cr
PS-264/17   &  264 & 17  &        &  3.5  $\pm$ 0.3 \cr\mr
PS-67/67    &  67  & 67  & 20.0   &  1.4  $\pm$ 0.2\cr
PS-264/67   &  264 & 67  & 20.0   &  1.4  $\pm$ 0.2 \cr\mr
PS-264/264  &  264 & 264 & 6.5    &  0.54 $\pm$ 0.12\cr
\br
\end{tabular}
\end{indented}
\end{table}
%
%

\section{Fluorescence correlation spectroscopy}
FCS is a method relying on the detection and temporal analysis of the 
fluorescence signal emitted from a small confocal detection volume 
(see e.g., refs. \cite{magd:72,rigl:01}). A laser beam is focused by an 
objective with high numerical aperture and excites fluorescent molecules 
entering the illuminated observation volume. Our FCS setup is based on the 
commercial ConfoCor2 setup (Carl Zeiss, Jena, Germany) \cite{rigl:01} with 
a 40$\times$ Plan Neofluar objective characterized by the numerical aperture 
NA=0.9. Fluorescence is excited by a He Ne-Ion laser at a wavelength of 
543~nm. For details of the FCS-measurements see 
refs. \cite{zett:04,zett:07,rigl:01,liu:05}.

The emitted fluorescent light is detected by an avalanche photo diode. The time 
dependent intensity fluctuations are analyzed by an autocorrelation function 
$G(\tau,c)$, where $\tau$ denotes the time. In order to account for 
the possibility of the contribution of both self-diffusion and cooperative 
diffusion, the function 
\begin{eqnarray} \label{eq1}
G(\tau,c)&=&\sum_{i \in \{s,co\}} G_i(0,c) 
\left(1+ \frac{4 D_i(c) \tau}{w_{x,y}^2}\right)^{-1}
\left(1+ \frac{4 D_i(c) \tau}{w_{z}^2}\right)^{-1/2}
\end{eqnarray} 
is used to describe the experimental data within a coupled-mode model 
(see ref. \cite{zett:09} and references therein). Here $w_{x,y}=296$ nm 
is the dimension of the observation volume perpendicular to the optical axis and  
$w_z=8 w_{x,y}$ is the dimension along the optical axis. $G_i(0,c)$ characterizes 
the contribution of the \mbox{$i$-th} component to the total amplitude $G(0,c)$ of the 
autocorrelation function. In \mbox{equation (\ref{eq1})} normal diffusion with mean 
square displacements $\phi_i(\tau,c)= 6 D_i(c) \tau$ has been assumed. Deviations 
from normal diffusion are due to internal chain motions (see e.g., refs.  
\cite{zett:09,harn:96,harn:98,harn:99,wink:06,shus:08} and references therein) or due 
to molecular crowding (see e.g., refs. 
\cite{zett:09,wach:00,weis:04,masu:05,bank:05,reit:08,cher:09} and references therein).

FCS is not only sensitive to intensity fluctuations due to the motion of labeled 
molecules but also due to photokinetic processes of the fluorescent dyes which 
occur for short times $\tau < 5\times 10^{-3}$ ms. This additional relaxation 
has been taken into account as discussed in refs. \cite{zett:04,zett:05,zett:07}.

\section{Autocorrelation functions measured by FCS}
\begin{figure}[ht!]
\begin{center}
\includegraphics[width=6.0cm,clip]{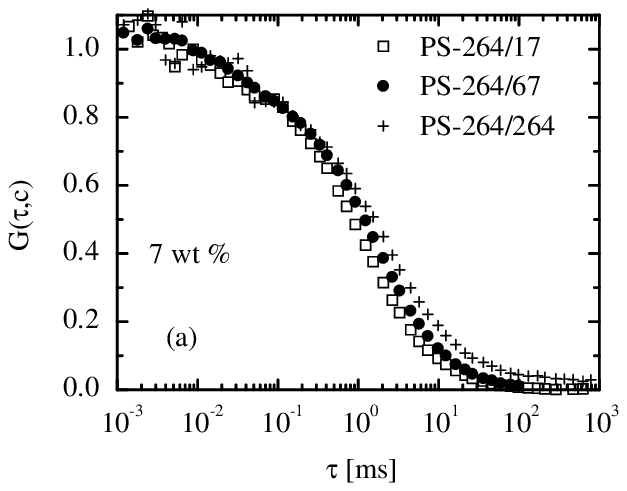}
\hspace*{0.25cm}
\includegraphics[width=6.0cm,clip]{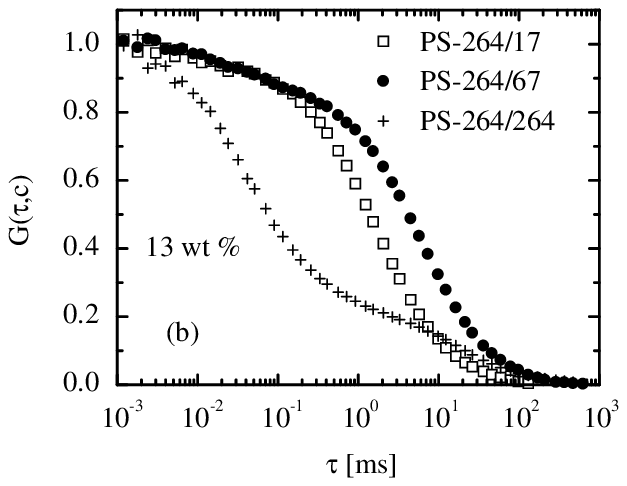}\\
\vspace*{0.5cm}
\includegraphics[width=6.0cm,clip]{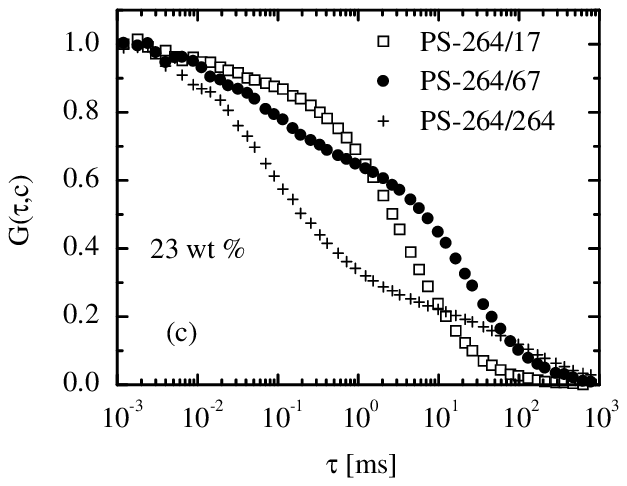}
\hspace*{0.25cm}
\includegraphics[width=6.0cm,clip]{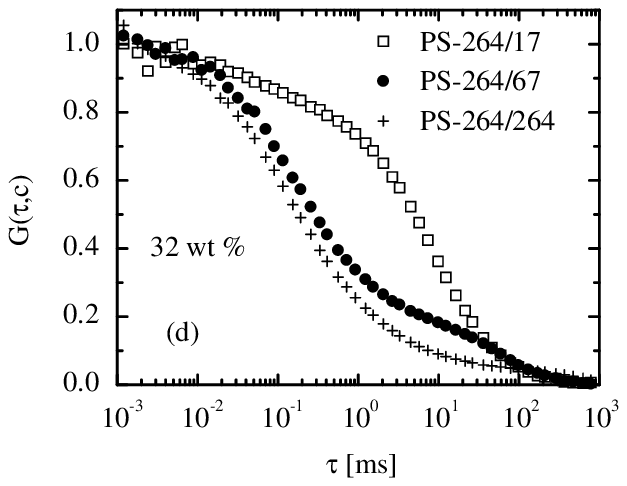}
\caption{Normalized autocorrelation functions obtained from FCS for the three 
samples PS-264/17, PS-264/67, and PS-264/264 (see table \ref{tab1}). The concentration 
of the matrix PS chains increases from (a) to (d) according to 7, 13, 23, 32 wt\%.}
\label{fig2}
\end{center}
\end{figure}

Two different types of macromolecular tracer diffusion behavior were obtained depending 
on the concentration and the molecular weight of the matrix polymer chains, as 
illustrated in figure \ref{fig2}. For the PS-264/17 sample containing short 
matrix PS chains ($M_{w}^{(ma)} = 17$ kg/mol, see table \ref{tab1}), the measured 
autocorrelation function is characterized by a single self-diffusion process 
for arbitrary concentrations (open squares in figures \ref{fig2} (a) - (d)).
With increasing concentration the decay of the 
autocorrelation function shifts to longer times. The simple self-diffusion model 
given by equation (\ref{eq1}) with $G_s(0,c)=1$ and $G_{co}(0,c)=0$ is successful in 
describing the autocorrelation data for such systems. This holds also for solutions 
containing longer matrix PS chains for low concentrations (PS-264/67 in figures \ref{fig2} 
(a), (b) and PS-264/264 in figure \ref{fig2} (a)). However, for higher concentrations, 
the simple self-diffusion model no longer describes the data in the case of the 
longer matrix PS chains. The autocorrelation data exhibit a second decay time on a 
shorter time scale (solid circles in figures \ref{fig2} (c), (d) and crosses 
in figures \ref{fig2} (b) - (d)). The experimental data can be described taking 
into account both self-diffusion and cooperative diffusion in equation (\ref{eq1}), 
i.e., $G_{co}(0,c)\neq 0$. The concentration $c^+$ at which the second 
fast diffusion process is detected depends on the molecular weight of the matrix PS 
chains. Upon decreasing the molecular weight of the matrix PS chains the 
concentration $c^+$ decreases (see table \ref{tab1}). In the case of the short 
matrix PS chains of molecular weight $M_{w}^{(ma)} = 17$ kg/mol no second diffusion 
process has been observed as already mentioned above. As the entanglement molecular 
weight of PS is 18 kg/mol, entanglements of matrix polymer chains are not 
possible at any concentration of the short matrix PS chains of molecular 
weight 17 kg/mol \cite{meer:85}.

An important result of the 
analysis of the autocorrelation functions for various concentrations is that the 
concentration $c^+$ is {\it independent} of the molecular weight of the tracer 
PS chains. For example, $c^+=20.0\, \mathrm{wt\%}$ for both the PS-67/67 and the 
PS-264/67 sample (see table \ref{tab1}). Moreover, only self-diffusion can 
be observed for both the PS-17/17 and the PS-264/17 sample, i.e., no concentration 
$c^+$ can be defined for these samples (see table \ref{tab1}). Hence the fast 
diffusion process is not a characteristic property of the tracer PS chains but is 
related to the dynamics of the surrounding matrix PS chains as will be discussed in 
more detail in the next section.
%
%
%

\section{Langevin and integral equation theory}
In this section we outline some points on the basic description of the 
cooperative dynamics of polymer solutions in terms of a Langevin and integral 
equation theory. The application of this approach to PS chains in solution 
has been discussed earlier \cite{zett:09,schw:97,harn:01}. Therefore, we only 
present the equations necessary for this study. The central assumption is that 
the total dynamic scattering function $S_{tot}(q,c,\tau)$ is a good variable 
for characterizing the cooperative dynamics of the system, where $q$ is the 
absolute value of the scattering vector. The time evolution of $S_{tot}(q,c,\tau)$ 
is assumed to be governed by a Langevin equation \cite{doi:86}. The total 
dynamic scattering function can be measured by dynamic light 
scattering (DLS). The cooperative diffusion coefficient $D_{co}(c)$ determines 
the decay rate of the total dynamic scattering function and is given by 
\begin{eqnarray}  \label{eq2}
D_{co}(c)&\stackrel{q \to 0}{=}&\frac{k_BT}{4\pi^2 \eta}\int\limits_0^\infty
d q_1\,\frac{S_{tot}(q_1,c,0) q_1^2}{S_{tot}(q,c,0) q^2}
\left(\frac{q_1^2+q^2}{2q_1q}
\log\left|\frac{q_1+q}{q_1-q}\right|-1\right)\,,
\end{eqnarray}  
where the temperature $T$ and the viscosity $\eta$ characterize the solvent. Furthermore, 
the total static scattering function reads 
\begin{equation}  \label{eq3}
S_{tot}(q,c,0)=1+\bar{v} c h(q,c)/(V_p P(q,c))\,,
\end{equation}
where $V_p$ is the volume of a dissolved polymer chain, $h(q,c)$
is a particle-averaged total correlation function, and \mbox{$\bar{v}=0.916$ cm$^3$/g} 
is the specific weight of PS \cite{schul:57}. The particle-averaged intramolecular
correlation function $P(q,c)$ characterizes the geometric shape of the polymer chains at a 
given concentration c. The overall size of the polymer chains is reduced considerably 
upon increasing the concentration implying a concentration dependence of the
particle-averaged intramolecular correlation function. Therefore, we consider 
the following particle-averaged intramolecular correlation function \cite{fuch:97}
\begin{equation}  \label{eq4}
P(q,c)=\left(1+0.549\, q^2 r_g^2(c)\right)^{-5/6}
\end{equation}
with the concentration dependent radius of gyration 
\begin{eqnarray}  \label{eq5}
r_g^2(c)&=& r_g^2(0) \left(\Theta(c^\star-c)+
\left(\frac{\displaystyle c}{\displaystyle c^*}\right)^{-1/8}\Theta(c-c^\star)\right)\,.
\end{eqnarray}
Here $\Theta(x)$ is the Heaviside step function which is 1 for $x>0$ and zero 
elsewhere. Moreover, the overlap concentration $c^\star$ is the boundary concentration 
between the dilute and semidilute regimes (see table \ref{tab1}). This overlap
concentration depends on the molecular weight according to 
$c^\star \sim (M_{w}^{(ma)})^{-4/5}$,
and has been determined for the PS solutions under considerations using FCS 
\cite{zett:07}. In addition the scaling law given $c^{-1/8}$ in equation (\ref{eq5}) 
has been confirmed experimentally for PS in a good solvent using small angle neutron 
scattering \cite{daou:75}. The particle-averaged total correlation function is related 
to a particle-averaged direct correlation function $C(q,c)$ by the generalized 
Ornstein-Zernike equation of the Polymer Reference Interaction Site Model (PRISM), 
which reads (see e.g., refs \cite{schw:97,harn:08} and references therein) 
\begin{equation}  \label{eq6}
h(q,c)=P^2(q,c)C(q,c)/(1-\bar{v} c C(q,c)P(q,c)/V_p)\,.
\end{equation}
This generalized Ornstein-Zernike equation is supplemented by the Percus-Yevick 
approximation to account for steric effects \cite{schw:97}. The osmotic pressure 
$p(c)$ is evaluated from equations (\ref{eq3}) - (\ref{eq6}) as
\begin{equation}  \label{eq7}
p(c)=k_BT \bar{v}\int\limits_0^c dc'\, S_{tot}(q,c',0)/V_p\,.
\end{equation}
The PRISM integral equation theory has been successfully applied to various 
polymer solutions (see e.g., refs \cite{schw:97,harn:01,boli:07,boli:09}).

\begin{figure}[ht!]
\begin{center}
\includegraphics[width=6.0cm, clip]{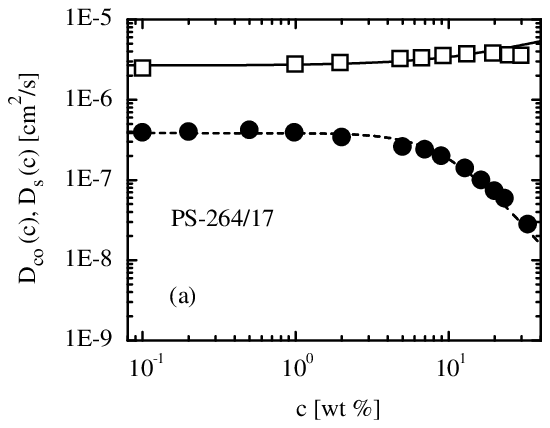}
\hspace*{0.25cm}
\includegraphics[width=6.0cm, clip]{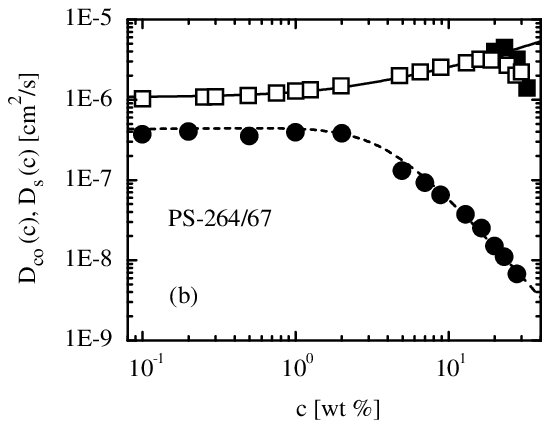}\\
\vspace*{0.5cm}
\includegraphics[width=6.0cm, clip]{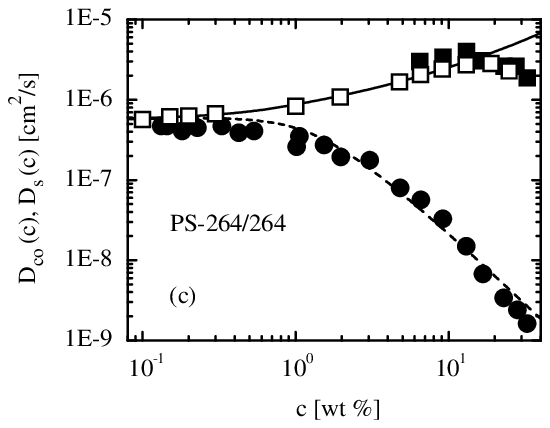}
\caption{Cooperative diffusion coefficients ($D_{co}(c)$, squares) and self-diffusion 
coefficients ($D_s(c)$, circles) for the three samples PS-264/17, PS-264/67, and 
PS-264/264 (see table \ref{tab1}) in panel (a), (b), and (c), respectively. Solid 
symbols refer to FCS measurements (see figure \ref{fig2}). The open squares denote 
DLS data obtained from the samples PS-17/17, PS-67/67, and PS-264/264 \cite{zett:09}. 
The solid lines display the collective diffusion coefficients as obtained from the 
Langevin and integral equation theory according to equations (\ref{eq2}) - (\ref{eq6}).
Dashed lines represent the calculated results as obtained from equation (\ref{eq8}) 
with equations (\ref{eq2}) and (\ref{eq7}) as input and $\alpha= 7.2, 2.5, 1$ in panel 
(a), (b), and (c), respectively.}
\label{fig3}
\end{center}
\end{figure}

Figures \ref{fig3} (a), (b), and (c) display the calculated cooperative diffusion 
coefficients of the 17, 67, and 264 kg/mol matrix PS chains (solid lines) together 
with the experimental data measured with DLS (open squares) \cite{zett:09} and FCS 
(solid squares). The figures demonstrate that the measured cooperative diffusion 
coefficients agree with the calculated results as obtained from equations 
(\ref{eq2}) - (\ref{eq6}). In particular, the crossover region between the dilute 
and the semidilute regimes is captured correctly by the Langevin and integral 
equation theory. The maximum of the cooperativ diffusion coefficient in the 
semidilute entangled regime marks the 
onset of glassy dynamics. This friction controlled dynamics is not captured by 
equation (\ref{eq2}). Therefore, deviations between the solid lines and the symbols 
are found for high concentrations in figure \ref{fig3}. Nevertheless, 
one can conclude from figures \ref{fig3} (b) and (c) that the diffusion coefficient 
$D_{co}(c)$ as obtained by FCS (solid squares) is indeed the cooperative diffusion 
coefficient of the matrix PS chains. The topological interactions in the semidilute 
solutions lead to coherent movements of matrix and tracer PS chains characterized by the 
cooperative diffusion coefficient $D_{co}(c)$. The resulting temporal fluctuations of the 
detected fluorescence intensity can be measured by FCS even in the case that the number 
of labeled tracer PS chains $N_l$ is {\it considerably} smaller than the number of 
matrix PS chains $N_{ma}$, i.e., $N_l \approx N_{ma} \times  10^{-6}$ for the 
systems under consideration.
%
%
%
%

The self-diffusion coefficient $D_s(c)$ as obtained using FCS measurements (see figure 
\ref{fig2}) and equation (\ref{eq1}) are also shown in figure \ref{fig3} (solid circles).
$D_s(c)$ is found to decrease with increasing concentration of the matrix PS chains 
due to the friction between the polymer chains. The dashed lines in figure \ref{fig3} 
are theoretical values calculated according to \cite{kane:05}
\begin{equation}  \label{eq8}
\frac{D_{co}(c)}{D_s(c)} = \alpha \left(1-\bar{v} c\right) \frac{d p(c)}{d c}
\end{equation}
with both $D_{co}(c)$ and $p(c)$ obtained from the Langevin and integral equation theory 
given by equations (\ref{eq2}) - (\ref{eq6}). As a new feature of the present evaluation, 
we have introduced the parameter $\alpha$ in equation (\ref{eq8}). This parameter describes 
the difference of the system under consideration from a  solution consisting 
of matrix polymer chains and tracer chains of the same molecular weight, that is, 
$\alpha=1$. The dashed line in figure \ref{fig3} (c) demonstrates that the measured 
self-diffusion coefficient of the  solution PS-264/264 can be described 
by equation (\ref{eq8}) with $\alpha=1$ and the Langevin and integral equation theory 
as input. Similarly, $D_s(c)$ can be calculated in agreement with experimental data 
for the samples PS-17/17 and PS-67/67 using $\alpha=1$ (data not shown). In order 
to describe the self-diffusion coefficients of the PS-264/67 and PS-264/17 samples, 
values of $\alpha=2.5$ and $\alpha=7.2$ above unity had to be chosen (dashed lines 
in figures \ref{fig3} (b) and (a). The values of $\alpha$ different from unity reflect 
the fact that the molecular weights of the matrix and tracer polymer chains are different
in the case of the samples PS-264/67 and PS-264/17.

Scaling arguments for self-avoiding random coils lead to the prediction 
$\alpha=(M_w^{(ma)}/M_w^{(tr)})^{-3/5}$, where the Flory exponent $\nu=3/5$ for a 
good solvent has been used. Hence one obtains $\alpha=2.3$ and $\alpha=5.2$ for the 
samples PS-264/67 and PS-264/17, respectively. The predicted value  $\alpha=2.3$ is 
close to the value  $\alpha=2.5$ used in our analysis in the case of the PS-264/67 
sample. This agreement confirms our earlier finding that the self-diffusion coefficients 
of both 264 kg/mol PS chains and 67 kg/mol PS chains fulfil scaling relations 
\cite{zett:09}. However, the 17 kg/mol PS chains are too short to be considered as 
self-avoiding random coils. Molecular stiffness leads to a more pronounced dependence 
of dynamical properties on the molecular weight than in the case of self-avoiding 
random coils \cite{harn:96,harn:98,harn:99}. Therefore, the value $\alpha=7.2$ used 
in our analysis is larger than  $\alpha=5.2$.

\begin{figure}[ht!]
\begin{center}
\includegraphics[width=6.0cm, clip]{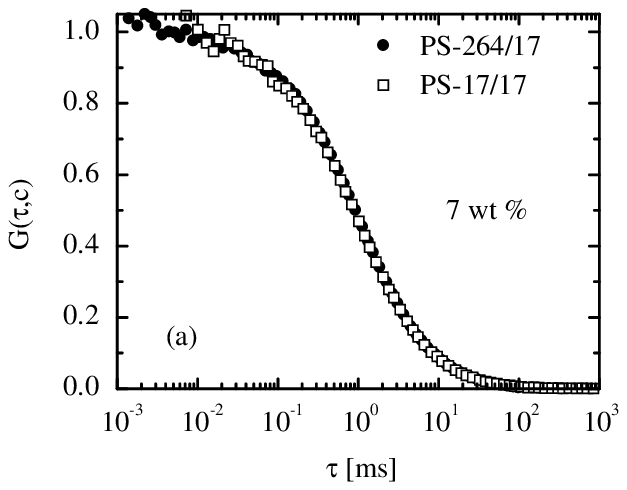}
\hspace*{0.25cm}
\includegraphics[width=6.0cm, clip]{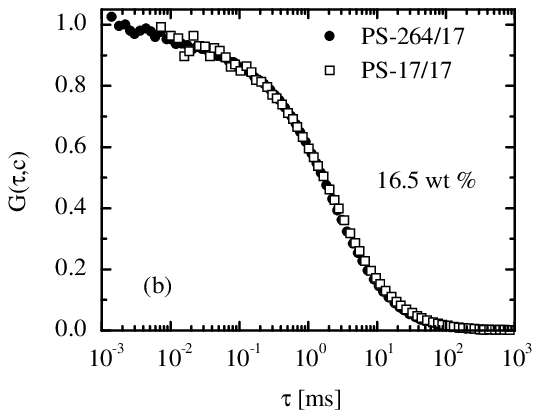}
\caption{Normalized FCS autocorrelation functions for the samples PS-264/17 and 
PS-17/17 (see table \ref{tab1}). The concentration of the 17 kg/mol matrix PS chains 
is 7 wt \% in panel (a) and 16.5 wt \% in panel (b). The autocorrelation functions 
of both samples coincide for a given concentration because the time $\tau$ for the 
PS-17/17 sample has been multiplied by the factor $\alpha=7.2$ as is discussed in 
the main text.}
\label{fig4}
\end{center}
\end{figure}

Figure \ref{fig4} demonstrates that the FCS autocorrelation functions 
for the PS-264/17 and PS-17/17 samples coincide provided the time is 
multiplied by the factor  $\alpha=7.2$ in the case of the PS-17/17 sample. This 
scaling is valid for all concentrations under consideration because the 17 kg/mol 
matrix PS chains do not form an entangled network in semidilute solution as mentioned 
earlier. Hence the polymeric nature of these short matrix chains does not lead to 
additional characteristic features of the FCS autocorrelation functions. In the 
case of a similar comparison of the FCS autocorrelation functions for the PS-264/67 
and PS-67/67 samples, scaling can be found only for concentrations lower than 
$c^+$. For higher concentrations deviations from a simple scaling law are 
found because the self-diffusion coefficient and the cooperative diffusion coefficient 
exhibit different dependencies on the molecular weight \cite{zett:09}.
%
%
%

\section{Conclusions} 
Fluorescence correlation spectroscopy has been used to study the dynamics of labeled 
tracer polystyrene chains in a system consisting of matrix polystyrene chains dissolved 
in toluene [figure \ref{fig1}]. The self-diffusion coefficient of the tracer polystyrene
chains has been measured for arbitrary concentrations of the matrix polystyrene chains.
Moreover, the cooperative diffusion coefficient has been determined in the semidilute
entangled concentration regime due to the transient entanglement network [figure \ref{fig2}]. The minimum concentration of matrix polystyrene chains at which the
cooperative diffusion coefficient can be detected by FCS is independent of the 
molecular weight of the tracer polystyrene chains [table \ref{tab1}]. It has been 
suggested earlier in the context of polymer fiber spinning that a polymer solution 
is converted to a more stable elastically deformable network at such a 
concentration \cite{schr:65,haya:67,mcke:04,shen:05,weit:08}.
Due to the resulting effective longe-range interaction of the polymer chains, the cooperative diffusion coefficient can be detected by FCS even in the case that the 
number of labeled polymer chains is considerably smaller than the number of unlabeled
polymer chains.  A theoretical description of the diffusion coefficients is given by 
a Langevin and integral equation theory [figure \ref{fig3}]. Moreover, a single 
master autocorrelation curve has been found for short unentangled polystyrene 
matrix chains [figure \ref{fig4}].

In summary, the present work gives further support to the recent conclusion that 
both the self-diffusion coefficient and the cooperative diffusion coefficient can 
be obtained experimentally using the same technique \cite{zett:09}.

\end{document}